\documentclass[a4paper,11pt]{article}
\pdfoutput=1 

\usepackage{jheppub} 

\usepackage[T1]{fontenc} 

\usepackage{amssymb}

\title{\boldmath $\rho-\omega$ mixing contribution to the measured $CP$ asymmetry of $B^\pm\to \omega K^\pm$}

\author[a]{Zhen-Hua Zhang}
\author[a]{Sheng Yang}
\author[b]{Xin-Heng Guo}

\affiliation[a]{School of Nuclear Science and Technology, University of South China, Hengyang, Hunan, 421001, China.}
\affiliation[b]{College of Nuclear Science and Technology, Beijing Normal University, Beijing, 100875, China.}

\emailAdd{zhangzh@usc.edu.cn}
\emailAdd{yangs\_usc@163.com}
\emailAdd{xhguo@bnu.edu.cn}

\abstract{
We study the $CP$ asymmetry of $B^\pm\to \omega K^\pm$
with the inclusion of the $\rho-\omega$ mixing mechanism.
It is shown that the $CP$ asymmetry of $B^\pm\to\omega K^\pm$ experimentally measured ($A_{CP}^{\text{exp}}$) and conventionally defined ($A_{CP}^{\text{con}}$) are in fact different, which relation can be illustrated as $A_{CP}^{\text{exp}}=A_{CP}^{\text{con}}+\Delta A_{CP}^{\rho\omega}$, with $\Delta A_{CP}^{\rho\omega}$ the $\rho-\omega$ mixing contribution to $A_{CP}^{\text{exp}}$.
The numerical value of $\Delta A_{CP}^{\rho\omega}$ is extracted from the experimental data of $B^\pm\to\pi^+\pi^-K^\pm$ and is found to be comparable with $A_{CP}^{\text{exp}}$, hence, non-negligible.
The conventionally defined $CP$ asymmetry, $A_{CP}^{\text{con}}$, is obtained from the values of $A_{CP}^{\text{exp}}$ and $\Delta A_{CP}^{\rho\omega}$, and is compared with the theoretical calculations in the literature.
}

\begin{document}
\maketitle
\flushbottom


\section{Introduction}
\label{sec.intro}

It was proposed long time ago that the mixing of the $\rho^0(770)$ and $\omega(782)$ resonances, which is termed as $\rho-\omega$ mixing, can affect $CP$ asymmetries of $B$ meson decays such as $B^\pm\to\pi^\pm\pi^+\pi^-$ and $B^\pm\to\rho^\pm\pi^+\pi^-$ \cite {Gardner:1997yx,Guo:2000uc,Wang:2015ula}, where $\rho^0\to \pi^+\pi^-$ is polluted by $\omega\to\rho^0\to\pi^+\pi^-$.
However, no definite confirmation of the contributions of the $\rho-\omega$ mixing effect on $CP$ asymmetries has ever been provided experimentally, although some hints on such contributions showed up in a recent analysis in Ref. \cite{Aaij:2019hzr}.
On the contrary, it is usually believed that $\rho-\omega$ mixing is hardly relevant for $B$ meson decays when the $\omega$ meson is involved in the final state, such as $B\to \omega K$ and $B \to \omega \pi$, as the contributions of $\rho^0\to\omega$ ($B\to \rho^0 K/\pi \to\omega K/\pi$) to the amplitudes are negligible.
This is a good approximation for branching ratios.
However, for $CP$ asymmetries, as will be seen, things may change.

For the $CP$ asymmetry of $B^\pm\to\omega K^\pm$, there seems to be a puzzle confronting the theoretical predictions and the experimental data.
On one hand, this $CP$ asymmetry has been studied extensively on the theoretical side via QCD factorization \cite{Cheng:2009cn}, perturbative QCD factorization \cite{Li:2006jv}, and Soft Collinear Effective Theory \cite{Wang:2008rk},
with the values $0.221^{+0.137+0.140}_{-0.128-0.130}$, $0.32^{+0.15}_{-0.17}$, and $0.116^{+0.182+0.011}_{-0.204-0.011}$ ($0.123^{+0.166+0.080}_{-0.173-0.011}$), respectively.\footnote{Other approaches such as the flavor diagram approach \cite{Cheng:2014rfa}, and the factorization assisted topological amplitude approach \cite{Zhou:2016jkv}, give predictions of $0.010\pm0.080$, and $0.19\pm0.09$, respectively.
Note however, these are in fact not pure theoretical approaches, since other data are used as inputs in these approaches.
}
Although with large uncertainties, these theoretical approaches tend to give a sizable $CP$ asymmetry in this decay channel.
On the other hand, the experimental value of this $CP$ asymmetry, which was measured by Belle and BaBar \cite{Aubert:2007si,Chobanova:2013ddr}, is numerically small and consistent with zero, with the latest world average \cite{Tanabashi:2018oca}
\begin{equation}\label{ACPexpnum}
A_{CP}^{\text{exp}}=-0.02\pm0.04.
\end{equation}
Currently, no satisfactory solution to this puzzle has been given in these theoretical approaches.

In this paper, we have no intention to solve the aforementioned puzzle.
Instead, in view of these troublesome theoretical predictions for the $CP$ asymmetry of $B^\pm\to\omega K^\pm$, we will estimate the contributions of the $\rho-\omega$ mixing effect to the $CP$ asymmetry of $B^\pm\to\omega K^\pm$, without utilizing any of these theoretical approaches.
To our best knowledge, this is the first work to study the  contributions of the $\rho-\omega$ mixing effect to $CP$ asymmetries of $B$ meson decays with an $\omega$ meson involved in the final state.
According to our analysis, this effect turns out to give an important contribution to the measured $CP$ asymmetry of $B^\pm\to\omega K^\pm$.

This paper is organized as follows.
In Sec. \ref{sec.form}, the $\rho-\omega$ mixing contribution to the $CP$ asymmetry of $B^\pm\to\omega K^\pm$ is analysed and separated.
In Sec. \ref{sec.estimation}, with the experimentally extracted decay amplitudes by BaBar, the $\rho-\omega$ mixing contribution to the $CP$ asymmetry of $B^\pm\to\omega K^\pm$ is estimated, and is found non-negligible.
Conclusion is made in Sec. \ref{sec.concl}.

\section{\boldmath $\rho-\omega$ mixing contributions to $CP$ asymmetry of $B^\pm\to\omega K^\pm$}
\label{sec.form}
The difference between the patterns of the $\rho-\omega$ mixing mechanism entering in $B^\pm\to\omega K^\pm$ and that in the previously studied channels such as $B^\pm\to\pi^\pm\pi^+\pi^-$ and $B^\pm\to\rho^\pm\pi^+\pi^-$ is apparent.
In the latter situation, the weakly produced $\omega$ meson transforms to the $\rho^0$ meson, which then finally decays into the $\pi^+\pi^-$ pair,
while for the situation of $B^\pm\to\omega K^\pm$, the process happens inversely, i.e. the weakly produced $\rho^0$ resonance transforms to the $\omega$ meson.
Due to the different properties between the $\rho^0$ and $\omega$ mesons, especially the large difference between their decay widths, the study patterns should be different.

Since the $\omega$ meson is usually reconstructed through the decay channel $\omega \to \pi^+\pi^-\pi^0$, the $CP$ asymmetry of $B^\pm\to\omega K^\pm$ being measured, $A_{CP}^{\text{exp}}$, is in fact the regional one of $B^\pm\to \pi^+\pi^-\pi^0 K^\pm$ when the invariant mass of the three-pion system lies in the vicinity of the $\omega$ resonance after the background subtraction, which can be expressed accordingly as
\begin{equation}\label{ACPexp_def}
A_{CP}^{\text{exp}}
\equiv
\frac{\int_{(m_{\omega}-\Delta_\omega)^2}^{(m_{\omega}+\Delta_\omega)^2} \left[\int \left( |\mathcal{A}^{-}|^2-|\mathcal{A}^{+}|^2\right) ds_{+0}ds_{-0}\right]ds}{\int_{(m_{\omega}-\Delta_\omega)^2}^{(m_{\omega}+\Delta_\omega)^2} \left[\int \left( |\mathcal{A}^{-}|^2+|\mathcal{A}^{+}|^2\right) ds_{+0}ds_{-0}\right]ds},
\end{equation}
where $\mathcal{A}^{\mp}$ is the decay amplitude of $B^\mp\to\pi^+\pi^-\pi^0 K^\mp$, $s_{+0}$, $s_{-0}$, and $s$ are the invariant masses squared of the $\pi^+\pi^0$, $\pi^-\pi^0$, and $3\pi$ systems, respectively, $m_\omega$ is the mass of the $\omega$ meson, the integral with respect to $s$ is performed around the vicinity of $\omega$, which has been taken between $(m_{\omega}-\Delta_\omega)^2$ and $(m_{\omega}+\Delta_\omega)^2$, with the cut $\Delta_\omega$ being taken such that the line shape of $\omega$ is included in the integral interval.\footnote{
When reconstructing the omega meson experimentally, the cut $\Delta_\omega$ is chosen to maximize the signal-to-background ratio, where the cut $\Delta_\omega$ is determined by the convolution of the width of the omega meson and the momentum resolution of the detector.
Typically, $\Delta_\omega$ is
comparable with the decay width of the $\omega$ meson, $\Gamma_\omega$.
See, for example, a recent paper of ALICE collaboration \cite{Acharya:2020tjy}.
}

For the situations when the invariant mass of the $3\pi$ system lies in the region of the $\omega$ resonance, the decay amplitude of $B^\mp\to \pi^+\pi^-\pi^0 K^\mp$ is dominated by the cascade decay $B^\mp\to \omega(\to \pi^+\pi^-\pi^0)K^\mp$, and potentially by the decay via the $\rho-\omega$ mixing effect, $B^\mp\to  \rho^0(\to\omega\to\pi^+\pi^-\pi^0)K^\mp$.
Consequently, the decay amplitude in this region
 can be expressed as
\begin{equation}\label{DecayAmp}
\mathcal{A}^\mp=\sum_{\lambda}\left[\mathcal{A}_{B^\mp\to \omega K^\mp}^\lambda+\delta_{\rho\omega}(s)\mathcal{A}_{B^\mp\to \rho K^\mp}^\lambda\right]\cdot\frac{\mathcal{A}_{\omega\to3\pi}^\lambda }{s_\omega},
\end{equation}
where $\mathcal{A}_{B^\mp\to \omega K^\mp}^\lambda$, $\mathcal{A}_{B^\mp\to \rho K^\mp}^\lambda$, and $\mathcal{A}_{\omega\to3\pi}^\lambda$ are the decay amplitudes of $B^\mp\to \omega K^\mp$, $B^\mp\to \rho^0 K^\mp$, and $\omega\to3\pi$, respectively, $\lambda$ is the polarization index of the intermediate vector resonances $\omega$ or $\rho^0$, and $\delta_{\rho\omega}(s)\equiv\frac{\Pi_{\rho\omega}^*(s)}{s_\rho}$, with $\Pi_{\rho\omega}$ the $\rho-\omega$ mixing parameter and $s_V=s-m_V^2-im_V\Gamma_V$ $(V=\omega, \rho)$.

If one parameterizes the decay amplitudes of $B^\mp\to \omega K^\mp$ and $B^\mp\to \rho^0 K^\mp$ as
$\mathcal{A}_{B^\mp\to \omega K^\mp}^\lambda=\mathcal{F}_{B^\mp\to \omega K^\mp}  \epsilon^*(q,\lambda)\cdot p_B$
and
$\mathcal{A}_{B^\mp\to \rho K^\mp}^\lambda=\mathcal{F}_{B^\mp\to \rho K^\mp}  \epsilon^*(q,\lambda)\cdot p_B$, respectively,
with $p_B$ the momentum of the $B^\mp$ meson, $\epsilon$ the polarization vector of the $\omega$ or $\rho^0$ resonances, $\mathcal{F}_{B^\mp\to \omega K^\mp}$ and $\mathcal{F}_{B^\mp\to \rho K^\mp}$ the rest parts of these two amplitudes, respectively, then the decay amplitude in Eq. (\ref{DecayAmp}) can be cast into
\begin{equation}\label{DecayAmp2}
\mathcal{A}^\mp=\left[\mathcal{F}_{B^\mp\to \omega K^\mp}+\delta_{\rho\omega}(s)\mathcal{F}_{B^\mp\to \rho K^\mp}\right]\cdot\frac{\mathcal{X}_{\omega\to3\pi}}{s_\omega},
\end{equation}
where $\mathcal{X}_{\omega\to3\pi}\equiv\sum_\lambda \mathcal{A}_{\omega\to3\pi}^\lambda\epsilon^*(q,\lambda)\cdot p_B$.
Once Eq. (\ref{DecayAmp2}) is substituted into Eq.(\ref{ACPexp_def}),
the experimentally measured $CP$ asymmetry can be expressed as
\begin{equation}\label{ACPexp_rhoomega}
A_{CP}^{\text{exp}}\approx\frac{\left|\mathcal{F}_{B^-\to \omega K^-}+\hat{\delta}_{\rho\omega}\mathcal{F}_{B^-\to \rho K^-}\right|^2-\left|\mathcal{F}_{B^+\to \omega K^+}+\hat{\delta}_{\rho\omega}\mathcal{F}_{B^+\to \rho K^+}\right|^2}{\left|\mathcal{F}_{B^-\to \omega K^-}+\hat{\delta}_{\rho\omega}\mathcal{F}_{B^-\to \rho K^-}\right|^2+\left|\mathcal{F}_{B^+\to \omega K^+}+\hat{\delta}_{\rho\omega}\mathcal{F}_{B^+\to \rho K^+}\right|^2},
\end{equation}
where $\hat{\delta}_{\rho\omega}\equiv \delta_{\rho\omega}(m_\omega^2)$.
In deriving Eq. (\ref{ACPexp_rhoomega}), we have neglected all the smooth dependence on $s$ by replacing it simply by $m_\omega^2$, as the integral over the line shape of $\omega$ with respect to $s$ is performed in a narrow interval around $m_\omega^2$ and the factor $|1/s_\omega|^2$ in the integrate behaves like $|1/s_\omega|^2\sim\delta(s-m_\omega^2)$.
On the contrary, the $s$-dependence of $1/s_\omega$ survives because this dependence is sharp.
However, the corresponding integrals $\int 1/|s_{\omega}|^2 ds$ in the numerator and the denominator have been cancelled out.
This is in fact the so-called narrow width approximation, in which the width (or the branching ratio) of the cascade decay $B^\pm\to \omega K^\pm \to \pi^+\pi^-\pi^0 K^\pm$ is factorized into the product of those of the decays $B^\pm\to \omega K^\pm$ and $\omega\to\pi^+\pi^-\pi^0$, i.e.
$\Gamma_{B^\pm\to \omega K^\pm \to \pi^+\pi^-\pi^0 K^\pm}\propto \Gamma_{B^\pm\to \omega K^\pm}\Gamma_{\omega\to\pi^+\pi^-\pi^0}$, with $\Gamma_{B^\pm\to \omega K^\pm}\propto |\mathcal{F}_{B^\pm\to \omega K^\pm}+\hat{\delta}_{\rho\omega}\mathcal{F}_{B^\pm\to \rho K^\pm}|^2$.

Note that $\hat{\delta}_{\rho\omega}$ is numerically very small,\footnote{
The real and imaginary parts of the $\rho-\omega$ mixing parameter $\Pi_{\rho\omega}$ when $s=m_\rho^2$ is fitted to be
\cite{Wolfe:2010gf}
\begin{eqnarray}
  \Re \left(\Pi_{\rho\omega}(m_{\rho}^2)\right)&=&-4620\pm220_{\text{model}}\pm170_{\text{data}} ~\text{MeV}^2, \nonumber\\
  \Im \left(\Pi_{\rho\omega}(m_{\rho}^2)\right)&=& -6100\pm1800_{\text{model}}\pm1110_{\text{data}}~\text{MeV}^2. \nonumber
\end{eqnarray}
Neglecting the $s$-dependence of $\Pi_{\rho\omega}$, one can easily see that $\hat{\delta}_{\rho\omega}$ is numerically very small,
\begin{equation}
  \hat{\delta}_{\rho\omega}\approx(0.049\pm0.018) +(0.045\pm0.003)i. \nonumber
\end{equation}}
which is the main reason why $\rho-\omega$ mixing is negligible for branching ratios of $B$ meson decays with $\omega$ involved in the final states.
However, its contribution to the measured $CP$ asymmetry, $A_{CP}^{\text{exp}}$, is not negligible in spite of the smallness of $\hat{\delta}_{\rho\omega}$.
To see this, let us make a Taylor expansion of $A_{CP}^{\text{exp}}$ up to $\mathcal{O}(\hat{\delta}_{\rho\omega})$, which reads
\begin{equation}\label{ACPexpapprox}
A_{CP}^{\text{exp}}= A_{CP}^{\text{con}}+\Delta A_{CP}^{\rho\omega},
\end{equation}
where
\begin{eqnarray}\label{ACPcon_def}
A_{CP}^{\text{con}}=\frac{\left|\mathcal{F}_{B^-\to \omega K^-}\right|^2-\left|\mathcal{F}_{B^+\to \omega K^+}\right|^2}{\left|\mathcal{F}_{B^-\to \omega K^-}\right|^2+\left|\mathcal{F}_{B^+\to \omega K^+}\right|^2},
\end{eqnarray}
is the conventionally defined $CP$ asymmetry of $B^\pm\to \omega K^\pm$ in the literature, where the contribution of $\rho-\omega$ mixing is absent, and
\begin{equation}\label{DeltaACP2}
  \Delta A_{CP}^{\rho\omega} =  \left(1-{A_{CP}^{\text{con}}}^2\right)
  \times\Re\left[\left(\frac{\mathcal{F}_{B^-\to \rho K^-}}{\mathcal{F}_{B^-\to \omega K^-}}-\frac{\mathcal{F}_{B^+\to \rho K^+}}{\mathcal{F}_{B^+\to \omega K^+}}\right)\hat{\delta}_{\rho\omega}\right],
\end{equation}
measures the contribution of the $\rho-\omega$ mixing effect to $A_{CP}^{\text{exp}}$.
Eq. (\ref{ACPexpapprox}) represents the deference between the experimentally measured $CP$ asymmetry $A_{CP}^{\text{exp}}$ and the conventionally defined one $A_{CP}^{\text{con}}$.

A rough but insightful order estimation gives $\Delta A_{CP}^{\rho\omega}\sim \sqrt{{\overline{\mathcal {B}}_{B^\pm\to\rho K^\pm}}/{\overline{\mathcal{B}}_{B^\pm\to\omega K^\pm}}}\hat{\delta}_{\rho\omega}$, with $\overline{\mathcal {B}}_{B^\pm\to\rho K^\pm}$ and $\overline{\mathcal{B}}_{B^\pm\to\omega K^\pm}$ the $CP$-averaged branching ratios of $B^\pm\to\rho K^\pm$ and $B^\pm\to\omega K^\pm$, respectively,
from which one can see that $\Delta A_{CP}^{\rho\omega}$ and $\hat{\delta}_{\rho\omega}$ are of the same order, provided that no strong cancellation happens between the two amplitude ratios in Eq. (\ref{DeltaACP2}).
In fact, a strong cancellation between these two terms is very unlikely because the observed large $CP$ asymmetry in $B^\pm\to \rho^0 K^\pm$ \cite{Aubert:2008bj} already indicates a considerable difference between $\mathcal{F}_{B^-\to \rho K^-}$ and $\mathcal{F}_{B^+\to \rho K^+}$.
Since $\hat{\delta}_{\rho\omega}$ is numerically the same order as the central value of the latest experimentally measured $A_{CP}^{\text{exp}}$, it follows that the contribution of  $\Delta A_{CP}^{\rho\omega}$ to $A_{CP}^{\text{exp}}$ is not negligible.

\section{\boldmath Estimation of $\Delta A_{CP}^{\rho\omega}$ with the amplitudes extracted from $B^\pm\to\pi^+\pi^- K^\pm$}
\label{sec.estimation}

In order to estimate $\Delta A_{CP}^{\rho\omega}$, one needs the four amplitudes in Eq. (\ref{DeltaACP2}).
However, just as was mentioned in the introduction of this paper, the commonly used theoretical approaches to these amplitudes failed in giving satisfactory predictions for the $CP$ asymmetry of $B^\pm\to\omega K^\pm$, which indicates that the amplitudes of $B^\pm\to\omega K^\pm$ -- especially their phases -- calculated by these theoretical approaches are unreliable.
In view of this,  the four amplitudes in Eq. (\ref{DeltaACP2}) used to estimate $\Delta A_{CP}^{\rho\omega}$ in what follows will be those simultaneously extracted from the data in stead.
The decay channel $B^\pm\to\pi^+\pi^- K^\pm$ is currently the only process in the literature which can be used for the simultaneous extraction of the four amplitudes, while $B^\pm\to\pi^+\pi^-\pi^0 K^\pm$ is not a suitable channel.
This can be seen from the different forms of the amplitudes of these two decay channels.
First of all, the decay amplitude for $B^\pm\to\pi^+\pi^- K^\pm$ is approximated as
\begin{equation}
\mathcal{A}(B^\pm\to \pi^+\pi^- K^\pm)|_{s\sim m_{\omega}^2}\propto\left(\mathcal{F}_{B^\pm\to\rho K^\pm}+\mathcal{F}_{B^\pm\to\omega K^\pm}\frac{{\tilde \Pi}_{\rho\omega}}{s_{\omega}}\right)\frac{1}{s_{\rho}},
\end{equation}
 when the invariant mass squared of the two pions $s$ lies around $m_{\omega}^2$, where ${\tilde \Pi}_{\rho\omega}$ is the effective $\rho-\omega$ mixing parameter with the inclusion of the $\omega\to \pi^+\pi^-$ contribution.
In this region, the factor $1/s_\omega$ in the second term varies rapidly because of the smallness of $\Gamma_\omega$.
Combining with the first term, this will lead to an easy-to-observe $s$-dependent interference effect.
The amplitudes $\mathcal{F}_{B\to\rho K}$ and $\mathcal{F}_{B\to\omega K}$ can then be extracted by fitting the experimentally observed $s$-dependent differential decay width.
For the situation $B^\pm\to\pi^+\pi^-\pi^0 K^\pm$, on the other hand, the factor $1/s_{\rho}$ in the second term of Eq. (\ref{DecayAmp}) is almost a constant when $s$ varies in the vicinity of the $\omega$ meson, which makes it almost impossible to distinguish the contributions of these two terms in Eq. (\ref{DecayAmp}) experimentally.

The latest amplitude analysis of $B^\pm\to \pi^+\pi^- K^\pm$ is performed by BaBar \cite{Aubert:2008bj},
where, the fitted amplitudes, which are denoted as $c_j$ and ${\bar c}_j$ therein, representing the decay through channel $j$, i.e. $B^-\to j \to \pi^+\pi^-K^-$ and $B^+\to j \to \pi^+\pi^-K^+$, respectively.
For simplicity, we will use a different notation for the fitted amplitudes, $c^\pm_j$, for the channel $B^\pm\to j \to \pi^+\pi^-K^\pm$.
What we are interested in are the two decay channels $B^\pm\to\omega K^\pm\to \pi^+\pi^- K^\pm$ and $B^\pm\to\rho K^\pm\to \pi^+\pi^- K^\pm$.
Since these are two cascade decays, the decay amplitudes can be expressed as
\begin{equation}
  c_{\omega}^{\pm}\propto\mathcal{F}_{B^\pm\to \omega K^\pm} g_{\omega\pi\pi}^{\text{eff}},
\footnote{
Strictly speaking, the experimentally extracted amplitudes also contain the contributions of $\rho-\omega$ mixing, i.e. $c_{\omega}^{\pm}\propto(\mathcal{F}_{B^\pm\to \omega K^\pm}+{\hat \delta}_{\rho\omega}\mathcal{F}_{B^\pm\to \rho K^\pm})g_{\omega\pi\pi}^{\text{eff}}$.
This means that the factor in $\Delta A_{CP}^{\rho\omega}$ corresponding to these amplitudes is replaced by
$\frac{\mathcal{F}_{B^-\to \rho K^-}}{\mathcal{F}_{B^-\to \omega K^-}+{\hat \delta}_{\rho\omega}\mathcal{F}_{B^-\to \rho K-}}-\frac{\mathcal{F}_{B^+\to \rho K^+}}{\mathcal{F}_{B^+\to \omega K^+}+{\hat \delta}_{\rho\omega}\mathcal{F}_{B-\to \rho K-}}$.
This replacement is safe since the resulted difference is of $\mathcal{O}({\hat \delta}_{\rho\omega}^2)$.
}
\end{equation}
and
\begin{equation}
  c_{\rho}^{\pm}\propto\mathcal{F}_{B^\pm\to \rho K^\pm} g_{\rho\pi\pi},
\end{equation}
respectively, where $g_{\omega\pi\pi}^{\text{eff}}$ and $g_{\rho\pi\pi}$ are the coupling constants of $\omega\to\pi^+\pi^-$
\footnote{
The coupling $g_{\omega\pi\pi}^{\text{eff}}$ includes both the direct decay $\omega\to\pi^+\pi^-$ and the decay via $\rho-\omega$ mixing, $\omega\to\rho\to\pi^+\pi^-$.}
and $\rho^0\to\pi^+\pi^-$, respectively.
Then, the $\rho-\omega$ mixing effect contributing to $A_{CP}^{\text{exp}}$ can be expressed as
\begin{equation}
  \Delta A_{CP}^{\rho\omega}\approx\Re\left[\left(\frac{c_{\rho}^-}{c_{\omega}^-}- \frac{c_{\rho}^+}{c_{\omega}^+}\right)\cdot \frac{\tilde{\Pi}_{\rho\omega}(m_\omega^2)\Pi_{\rho\omega}^*(m_\omega^2)}{\left(m_\omega^2-m_\rho^2+im_\rho\Gamma_\rho\right)^2} \right],
\end{equation}
where we have made the replacement of $(1-{A_{CP}^{\text{con}}}^2)$ by 1, and used the relation
 $\tilde{\Pi}_{\rho\omega}(s)=s_\rho g_{\omega\pi\pi}^{\text{eff}}/g_{\rho\pi\pi}$
\footnote{
This relation can be understood as the following.
On one hand, the effective coupling $g_{\omega\pi\pi}^{\text{eff}}$ can be expressed as $g_{\omega\pi\pi}^{\text{eff}}=g_{\omega\pi\pi}+\frac{\Pi_{\rho\omega}(s)}{s_\rho}g_{\rho\pi\pi}$, where the first term represents the direct decay of $\omega$ to $\pi^+\pi^-$, while the second term represents the decay through $\rho-\omega$ mixing, i.e. $\omega$ first transforms to $\rho$, which decays into $\pi^+\pi^-$.
On the other, the effective $\rho-\omega$ mixing parameter ${\tilde \Pi}_{\rho\omega}$ is defined such that the direct decay $\omega\to\pi^+\pi^-$ is absorbed in it, i.e. $\frac{{\tilde \Pi}_{\rho\omega}(s)}{s_\rho}g_{\rho\pi\pi}=\frac{\Pi_{\rho\omega}(s)}{s_\rho}g_{\rho\pi\pi}+g_{\omega\pi\pi}$.
By comparing these two equations, one gets ${\tilde \Pi}_{\rho\omega}(s)=s_\rho \frac{g_{\omega\pi\pi}^{\text{eff}}}{g_{\rho\pi\pi}}$.
} with $s$ taken to be $m_\omega^2$.
According to Ref. \cite{Aubert:2008bj}, these amplitudes are parameterized as $c_{\omega(\rho)}^\pm=(x_{\omega(\rho)}\pm\Delta x_{\omega(\rho)})+i(y_{\omega(\rho)}\pm\Delta y_{\omega(\rho)})$, $\Delta A_{CP}^{\rho\omega}$ then transforms to
\begin{equation}
  \Delta A_{CP}^{\rho\omega}
  \approx-2\Re\left[\frac{\Delta x_\rho+i\Delta y_\rho}{x_{\omega}+iy_{\omega}} \cdot \frac{\tilde{\Pi}_{\rho\omega}(m_\omega^2)\Pi_{\rho\omega}^*(m_\omega^2)}{\left(m_\omega^2-m_\rho^2+im_\rho\Gamma_\rho\right)^2} \right]. \end{equation}
The factor corresponding the $\rho-\omega$ mixing effect can be extracted from the fitted parameters in Ref. \cite{Wolfe:2010gf}, which reads
\begin{equation}
  \frac{\tilde{\Pi}_{\rho\omega}(m_\omega^2)\Pi_{\rho\omega}^*(m_\omega^2)}{\left(m_\omega^2-m_\rho^2+im_\rho\Gamma_\rho\right)^2}=\left[(-2.20\pm0.33)+(1.30\pm0.46)i \right]\times10^{-3}.
\end{equation}
According to Ref. \cite{Aubert:2008bj}, $x_\omega=-0.058\pm0.067\pm0.018^{+0.053}_{-0.011}$, $y_\omega= 0.100\pm0.051\pm0.010^{+0.033}_{-0.031}$, $\Delta x_\omega=\Delta y_\omega=0$, $\Delta x_\rho=-0.160\pm0.049\pm0.024^{+0.094}_{-0.013}$,
and $\Delta y_\rho=0.169\pm0.096\pm0.057^{+0.133}_{-0.027}$.
Then, the $\rho-\omega$ mixing effect contributing to $A_{CP}^{\text{exp}}$ is extracted to be
\begin{equation}\label{DeltaACPnum}
  \Delta A_{CP}^{\rho\omega}  \approx 0.01^{+0.01}_{-0.02},
\end{equation}
where the uncertainty is estimated based on the extracted amplitudes by BaBar and ${\Pi}_{\rho\omega}$.
From a comparison of Eqs. (\ref{ACPexpnum}) and (\ref{DeltaACPnum}), it is concluded that the $\rho-\omega$ mixing effect is indeed not negligible.

Combining Eq. (\ref{DeltaACPnum}) with the experimental data for $A_{CP}^{\text{exp}}$, the conventionally defined $CP$ asymmetry of $B^\pm\to \omega K^\pm$ is obtained accordingly,
\begin{equation}\label{ACPthNum}
A_{CP}^{\text{con}} =-0.03\pm0.04 ^{+0.02}_{-0.01},
\end{equation}
where the first uncertainty comes from the world average $A_{CP}^{\text{exp}}$, and the second one is from $\Delta A_{CP}^{\rho\omega}$.
This number should be compared with the theoretical predictions of the $CP$ asymmetry of $B^\pm\to \omega K^\pm$ via different approaches.
Note however, one should not take the number in Eq. (\ref{ACPthNum}) too seriously, as the extracted amplitudes of BaBar in Ref. \cite{Aubert:2008bj} are not that reliable for our purpose for at least two reasons.
First of all, the $CP$ asymmetry of $B^\pm\to \omega K^\pm$ is set to be zero by hand in Ref. \cite{Aubert:2008bj}.
Secondly, the extracted amplitudes of $B^\pm\to\omega K^\pm$ suffer from large uncertainties, as their fractions in that of $B^\pm\to\pi^+\pi^- K^\pm$ are quite small.
Future extractions of the amplitudes with smaller uncertainties and more in line with our purpose are needed both for $A_{CP}^{\text{exp}}$ and $\Delta A_{CP}^{\rho\omega}$.
Of course, a theoretical calculation of $A_{CP}^{\text{exp}}$ with the explicit inclusion of the $\rho-\omega$ mixing effect is also desirable.

\section{Conclusion}
\label{sec.concl}
To sum up, it is concluded that the $CP$ asymmetry of $B^-\to \omega K^-$ being measured experimentally $A_{CP}^{\text{exp}}$ is in fact the regional one of $B^\pm\to\pi^+\pi^-\pi^0 K^\pm$ when the invariant mass of the $3\pi$ system lies around the $\omega$ resonance, which is different from the conventionally defined one, $A_{CP}^{\text{con}}$.
This conclusion is expressed as $A_{CP}^{\text{exp}}=A_{CP}^{\text{con}}+\Delta A_{CP}^{\rho\omega}$, where the $\rho-\omega$ mixing contribution, $\Delta A_{CP}^{\rho\omega}$, is not negligible.
This is supported by the amplitude analysis of the channel $B^\pm\to\pi^+\pi^- K^\pm$, from which $A_{CP}^{\text{con}}$ is also extracted and compared with the theoretical calculations.

The above analysis can be potentially generalized to other $B$ meson decays with the involvement of the $\omega$ meson as a final state particle.
For example, the similar pattern shows up in the decay $B^\pm\to\omega\pi^\pm$, where the $CP$ asymmetry is also comparable with $\hat{\delta}_{\rho\omega}$ \cite{Aubert:2007si,Aaij:2019hzr}.
In general, for the decay process $B\to \omega X$ ($X$ represents one or more particles) and its $CP$ conjugate $\overline{B}\to\omega \overline{X}$, the difference between the regional $CP$ asymmetry of $B\to \pi^+\pi^-\pi^0 X$ with the invariant mass of the three pions lying around the $\omega$ resonance and that of $B\to \omega X$ will be
\begin{equation}
  \Delta A_{CP}^{\rho\omega}(B\to \omega X)=\left(1-{A_{CP,B\to\omega X}^{\text{con}}}^2\right)\Re\left[\left(\frac{\mathcal{A}_{ \rho}}{\mathcal{A}_{\omega}}-\frac{\overline{\mathcal{A}}_{\rho}}{\overline{\mathcal{A}}_{\omega }}\right)\hat{\delta}_{\rho\omega}\right],
\end{equation}
where $\mathcal{A}_{V}$ and $\overline{\mathcal{A}}_V$ are the decay amplitudes of $B\to V X$ and $\overline{B}\to V \overline{X}$, respectively, ${A_{CP,B\to \omega X}^{\text{con}}}$ is the conventionally defined $CP$ asymmetry of $B\to\omega X$ without the contribution of $\rho-\omega$ mixing.
Strictly speaking, careful analysis of the $\rho-\omega$ mixing contributions should be performed for $CP$ asymmetries of most (if not all) $B$ meson decays when $\omega$ appears in the final states, since ${\hat \delta}_{\rho\omega}$ may give considerable contributions to the $CP$ asymmetries.

\acknowledgments
This work was supported by National Natural Science Foundation of China under Contracts Nos.11705081 and 11775024.

\bibliography{zzhbib}

\providecommand{\href}[2]{#2}\begingroup\raggedright\begin{thebibliography}{10}

\bibitem{Gardner:1997yx}
S.~Gardner, H.~B. O'Connell and A.~W. Thomas, \emph{{Rho - $\omega$ mixing and
  direct CP violation in hadronic $B$ decays}},
  \href{https://doi.org/10.1103/PhysRevLett.80.1834}{\emph{Phys. Rev. Lett.}
  {\bfseries 80} (1998) 1834}
  [\href{https://arxiv.org/abs/hep-ph/9705453}{{\ttfamily hep-ph/9705453}}].

\bibitem{Guo:2000uc}
X.-H. Guo, O.~M. Leitner and A.~W. Thomas, \emph{{Enhanced direct CP violation
  in B+- ---> rho0 pi+-}},
  \href{https://doi.org/10.1103/PhysRevD.63.056012}{\emph{Phys. Rev. D}
  {\bfseries 63} (2001) 056012}
  [\href{https://arxiv.org/abs/hep-ph/0009042}{{\ttfamily hep-ph/0009042}}].

\bibitem{Wang:2015ula}
C.~Wang, Z.-H. Zhang, Z.-Y. Wang and X.-H. Guo, \emph{{Localized direct CP
  violation in $B^\pm \rightarrow \rho ^0 (\omega )\pi ^\pm \rightarrow \pi ^+
  \pi ^-\pi ^\pm $}},
  \href{https://doi.org/10.1140/epjc/s10052-015-3757-2}{\emph{Eur. Phys. J. C}
  {\bfseries 75} (2015) 536}
  [\href{https://arxiv.org/abs/1506.00324}{{\ttfamily 1506.00324}}].

\bibitem{Aaij:2019hzr}
{\scshape LHCb} collaboration, \emph{{Observation of Several Sources of $CP$
  Violation in $B^+ \to \pi^+ \pi^+ \pi^-$ Decays}},
  \href{https://doi.org/10.1103/PhysRevLett.124.031801}{\emph{Phys. Rev. Lett.}
  {\bfseries 124} (2020) 031801}
  [\href{https://arxiv.org/abs/1909.05211}{{\ttfamily 1909.05211}}].

\bibitem{Cheng:2009cn}
H.-Y. Cheng and C.-K. Chua, \emph{{Revisiting Charmless Hadronic B(u,d) Decays
  in QCD Factorization}},
  \href{https://doi.org/10.1103/PhysRevD.80.114008}{\emph{Phys. Rev. D}
  {\bfseries 80} (2009) 114008}
  [\href{https://arxiv.org/abs/0909.5229}{{\ttfamily 0909.5229}}].

\bibitem{Li:2006jv}
H.-n. Li and S.~Mishima, \emph{{Penguin-dominated B ---> PV decays in NLO
  perturbative QCD}},
  \href{https://doi.org/10.1103/PhysRevD.74.094020}{\emph{Phys. Rev. D}
  {\bfseries 74} (2006) 094020}
  [\href{https://arxiv.org/abs/hep-ph/0608277}{{\ttfamily hep-ph/0608277}}].

\bibitem{Wang:2008rk}
W.~Wang, Y.-M. Wang, D.-S. Yang and C.-D. L{\"u}, \emph{{Charmless Two-body
  B(B(s)) ---> VP decays In Soft-Collinear-Effective-Theory}},
  \href{https://doi.org/10.1103/PhysRevD.78.034011}{\emph{Phys. Rev. D}
  {\bfseries 78} (2008) 034011}
  [\href{https://arxiv.org/abs/0801.3123}{{\ttfamily 0801.3123}}].

\bibitem{Cheng:2014rfa}
H.-Y. Cheng, C.-W. Chiang and A.-L. Kuo, \emph{{Updating B$\rightarrow$PP,VP
  decays in the framework of flavor symmetry}},
  \href{https://doi.org/10.1103/PhysRevD.91.014011}{\emph{Phys. Rev. D}
  {\bfseries 91} (2015) 014011}
  [\href{https://arxiv.org/abs/1409.5026}{{\ttfamily 1409.5026}}].

\bibitem{Zhou:2016jkv}
S.-H. Zhou, Q.-A. Zhang, W.-R. Lyu and C.-D. L{\" u}, \emph{{Analysis of
  Charmless Two-body B decays in Factorization Assisted Topological Amplitude
  Approach}}, \href{https://doi.org/10.1140/epjc/s10052-017-4685-0}{\emph{Eur.
  Phys. J. C} {\bfseries 77} (2017) 125}
  [\href{https://arxiv.org/abs/1608.02819}{{\ttfamily 1608.02819}}].

\bibitem{Aubert:2007si}
{\scshape BaBar} collaboration, \emph{{Branching fraction and CP-violation
  charge asymmetry measurements for B-meson decays to eta K+-, eta pi+-,
  eta-prime K, eta-prime pi+-, omega K, and omega pi+-}},
  \href{https://doi.org/10.1103/PhysRevD.76.031103}{\emph{Phys. Rev. D}
  {\bfseries 76} (2007) 031103}
  [\href{https://arxiv.org/abs/0706.3893}{{\ttfamily 0706.3893}}].

\bibitem{Chobanova:2013ddr}
{\scshape Belle} collaboration, \emph{{Measurement of branching fractions and
  CP violation parameters in $B\to\omega K$ decays with first evidence of CP
  violation in $B^0 \to \omega K^0_S$}},
  \href{https://doi.org/10.1103/PhysRevD.90.012002}{\emph{Phys. Rev. D}
  {\bfseries 90} (2014) 012002}
  [\href{https://arxiv.org/abs/1311.6666}{{\ttfamily 1311.6666}}].

\bibitem{Tanabashi:2018oca}
{\scshape Particle Data Group} collaboration, \emph{{Review of Particle
  Physics}}, \href{https://doi.org/10.1103/PhysRevD.98.030001}{\emph{Phys. Rev.
  D} {\bfseries 98} (2018) 030001}.

\bibitem{Acharya:2020tjy}
{\scshape ALICE} collaboration, \emph{{Production of $\omega$ mesons in pp
  collisions at $\sqrt{s}$ = 7 TeV}},
  \href{https://arxiv.org/abs/2007.02208}{{\ttfamily 2007.02208}}.

\bibitem{Wolfe:2010gf}
C.~Wolfe and K.~Maltman, \emph{{Consequences of the BaBar $e^+e^- \to pi^+pi^-$
  Measurement for the Determination of Model-Dependent $\rho-\omega$ Mixing
  Effects in $\Pi_{\rho\omega}(m_{\rho}^2)$ and $(g-2)_\mu$}},
  \href{https://doi.org/10.1103/PhysRevD.83.077301}{\emph{Phys. Rev. D}
  {\bfseries 83} (2011) 077301}
  [\href{https://arxiv.org/abs/1011.4511}{{\ttfamily 1011.4511}}].

\bibitem{Aubert:2008bj}
{\scshape BaBar} collaboration, \emph{{Evidence for Direct CP Violation from
  Dalitz-plot analysis of $B^\pm \to K^\pm \pi^\mp \pi^\pm$}},
  \href{https://doi.org/10.1103/PhysRevD.78.012004}{\emph{Phys. Rev. D}
  {\bfseries 78} (2008) 012004}
  [\href{https://arxiv.org/abs/0803.4451}{{\ttfamily 0803.4451}}].

\end{thebibliography}\endgroup

\end{document}